\documentclass[10pt,journal,compsoc]{IEEEtran}
\usepackage{comment}
\usepackage{xcolor}
\usepackage{subcaption}
\usepackage{authblk}
\usepackage{amsmath,amssymb,amsfonts}
\usepackage{mathtools}

\usepackage{algorithmic}
\usepackage{graphicx}
\usepackage{textcomp}
\usepackage{enumerate}

\usepackage{multicol}

\ifCLASSOPTIONcompsoc
  \usepackage[nocompress]{cite}
\else
  \usepackage{cite}
\fi
\ifCLASSINFOpdf
\else
\fi
\begin{document}
%
\title{Optimal energy-aware task scheduling for batteryless IoT devices}
%
%
%
%

\author[1,2]{Carmen Delgado\thanks{Corresponding author: carmen.delgado@i2cat.net}}
\author[1]{Jeroen Famaey}
\affil[1]{\textit{IDLab, University of Antwerp - imec}, Antwerp, Belgium}
\affil[2]{\textit{i2CAT Foundation}, Barcelona, Spain}

\markboth{IEEE Transactions on Emerging Topics in Computing, ~Vol.~, No.~, May~2021}%
{Shell \MakeLowercase{\textit{et al.}}: Bare Demo of IEEEtran.cls for Computer Society Journals}
%



\IEEEtitleabstractindextext{%
\begin{abstract}
Today's IoT devices rely on batteries, which offer stable energy storage but contain harmful chemicals. Having billions of IoT devices powered by batteries is not sustainable for the future. As an alternative, batteryless devices run on long-lived capacitors charged using energy harvesters. The small energy storage capacity of capacitors results in intermittent on-off behaviour. Traditional computing schedulers can not handle this intermittency, and in this paper we propose a first step towards an energy-aware task scheduler for constrained batteryless devices. We present a new energy-aware task scheduling algorithm that is able to optimally schedule application tasks to avoid power failures, and that will allow us to provide insights on the optimal look-ahead time for energy prediction. Our insights can be used as a basis for practical energy-aware scheduling and energy availability prediction algorithms. We formulate the scheduling problem as a Mixed Integer Linear Program. We evaluate its performance improvement when comparing it with state-of-the-art schedulers for batteryless IoT devices. Our results show that making the task scheduler energy aware avoids power failures and allows more tasks to successfully execute. Moreover, we conclude that a relatively short look-ahead energy prediction time of 8 future task executions is enough to achieve optimality.
\end{abstract}

\begin{IEEEkeywords}
Internet of Things (IoT); batteryless IoT devices; task scheduler; energy-aware; low-power wide-area networks (LPWAN); energy harvesting; optimization; Mixed Integer Linear Programming (MILP) 
\end{IEEEkeywords}}

\maketitle

\IEEEdisplaynontitleabstractindextext

%
\IEEEpeerreviewmaketitle

\IEEEraisesectionheading{\section{Introduction}\label{sec:introduction}}

%
%
%
%

\IEEEPARstart{T}{he} Internet of Things (IoT), where tens of billions of interconnected devices communicate and cooperate with each other over the Internet, is getting more and more attention nowadays. 
This is due to many reasons, but the most important ones are that these devices aim at supporting and improving daily life, they are cheap, and they are easy to use. 
Normally, these devices are equipped with a battery, a radio chip, a microcontroller unit (MCU) and one or more sensors and/or actuators. 
With the advancements in low-power and miniature electronics and in low power radio technologies, there has been a clear increase of IoT applications covering a wide range of application areas \cite{Ma2020}, such as home automation, wearable devices and industrial or agricultural monitoring. 

However, since their inception, batteries have been one of the main drivers of these IoT devices. But batteries are not only incompatible with a sustainable IoT since they contain harmful chemicals~\cite{Hester2017}, they are also sensitive to temperature changes, dangerous when not carefully protected, and short-lived, requiring costly maintenance and replacement every few years at best. Although rechargeable batteries in combination with energy harvesters can somehow offset this problem, they still suffer from capacity degradation due to frequent charge-discharge cycles, as well limiting their lifetime to a few years. This results in millions upon millions of discarded IoT batteries every year, filled with dangerous chemicals that significantly affect our environment and ecology. 
Moreover, batteries are susceptible to current peaks, which makes them degrade faster. Sadly, IoT devices often suffer from such peaks, due to the fact they spend most of their time in a low-power (sleep) state. When they wake up to transmit or receive data, their power consumption jumps up many orders of magnitude, resulting in short-lived current peaks. 
To address all these IoT-related battery problems, researchers have recently started investigating batteryless IoT devices and networks \cite{Muratkar2020}.

\begin{figure}[t]
\centerline{\includegraphics[width=1\columnwidth]{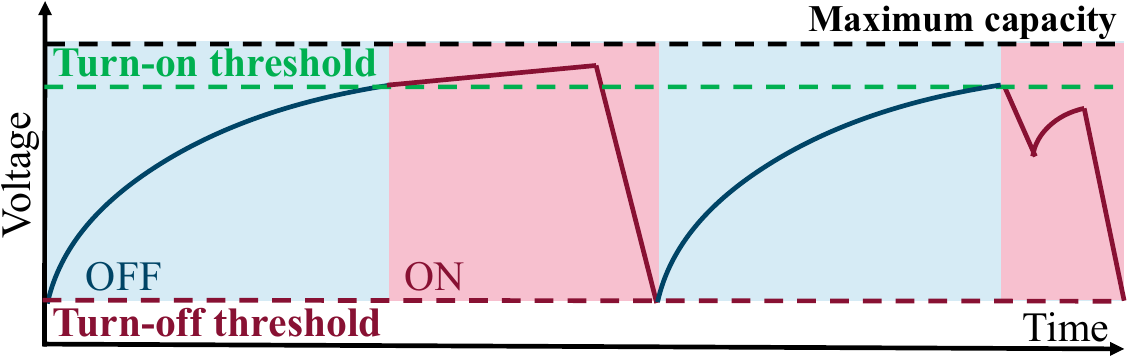}}
\caption{Batteryless device intermittent behaviour}
\label{fig:behaviour}
\end{figure}

These batteryless devices run on small but long-lived capacitors for energy storage, charged using various forms of energy harvesting (e.g., thermal, solar, vibration), which make them more environmentally friendly, cheaper to maintain, easy to recycle and more resistant to temperature variations and charge-discharge degradation. This makes them especially suitable for applications in hard-to-reach locations (e.g., intra-body health monitoring, remote-area sensing) and large-scale deployments (e.g., dense building automation networks, smart cities).
However, the combination of small energy storage capacities and stochastic energy harvesting behaviour causes batteryless devices to intermittently turn on and off due to frequent power failures (c.f Figure~\ref{fig:behaviour}). This results in a power failure when the capacitor voltage drops below the turn-off threshold. When the device harvests enough energy, it will turn on again when the turn-on voltage threshold is reached.

Such intermittency challenges the fundamental assumption that devices can operate uninterrupted to perform their tasks, and requires rethinking computing, communications and networking paradigms. In this paper we focus on the computing part, where we present an optimal energy-aware task scheduling algorithm for batteryless devices. 
We propose a Mixed Integer Linear Programming (MILP) optimization framework which intelligently decides when to execute which task, according to the energy harvested and stored, energy consumed by the tasks and their priority. 
We follow the same approach as in \cite{Yildirim2018}, where the programmer decomposes the application into a collection of interconnected atomic tasks. The runtime keeps track of the active task, re-executing it after a power failure, and keeping its output in non-volatile (i.e., permanent) memory after successful completion for use as input to other tasks. However, our approach is able to avoid power failures, improving the overall performance of the scheduler. 
The proposed optimal algorithm not only shows the potential maximum performance improvement that can be achieved with energy-aware scheduling compared to non-energy-aware scheduling, but it also offers insights in the look-ahead window in terms of predicting the future available energy that is needed to achieve such a gain. 
This means that thanks to our approach, designers will be able to define their energy harvesting prediction windows, as well as expected tasks to be executed. 
Although the presented MILP formulation cannot be directly solved in real-time on a batteryless IoT device, the proposed scheduling formulation and obtained insights can be used as a basis for the design of fast heuristic scheduling algorithms that can be executed in real-time on batteryless devices.

The remainder of this paper is organized as follows. Section~\ref{sec:sota} provides an overview of the related work in the field of batteryless computing and tasks schedulers. Section~\ref{sec:model} describes the proposed system model. In Section~\ref{sec:optimization}, we introduce our optimization framework.
The evaluation of the optimal task scheduling is provided in Section~\ref{sec:evaluation}. Finally, conclusions and future work are discussed in Section~\ref{sec:conclusions}.

\section{Related work}\label{sec:sota}

The batteryless vision requires computing mechanisms to deal with this intermittent behaviour. Traditional sequential computing models and programming languages cannot handle such abrupt intermittent behavior, as they assume uninterrupted execution of programming instructions and rely on volatile memory to keep application execution progress. 
Different computing models and schedulers to overcome the intermittent execution in batteryless devices have already been proposed. 
In their recent work, Sandhu \textit{et al.} survey existing task scheduling schemes for energy harvesting IoT, which aim at ensuring optimal utilization of harvested energy to extend system lifetime as well to provide highest activity detection/monitoring performance \cite{Sandhu2020}. They analyze the three major strategies employed: Dynamic Voltage and Frequency Scaling (DVFS), decomposing and combining of tasks and duty cycling. First, DVFS adjusts the voltage level and the frequency to power the active hardware module. Second, large tasks can be decomposed into smaller atomic subtasks. Finally, duty cycling is another task scheduling mechanism that allows controlling the consumed energy by the nodes when they are not performing any useful operation. DVFS algorithms are difficult to implement on these energy-constrained sensor nodes, due to the stringent requirement of complex circuitry that provides various voltage levels for different components of the node \cite{Sandhu2020}. For this reason, in this work we propose a new optimization algorithm that uses the other two strategies: we divide tasks into atomic subtasks and we also use duty cycling in order to let the device intelligently sleep in order to harvest more energy that will benefit the execution of future tasks.

The two main computing strategies are checkpointing- and task-based models, where these last ones are based on the task decomposing strategy described above.
While checkpointing-based models such as Mementos \cite{Ransford2011} and Clank \cite{Hicks2017} are not scalable due to the time and energy cost to create checkpoints, which increases with the size of the volatile memory, task-based models are more suitable for batteryless devices. As mentioned before, these models divide the program into different atomic subtasks. The output of a task is stored in non-volatile memory when it successfully completes.
To reduce the overhead, task-based models have been proposed, dividing the program into different atomic subtasks. The output of a task is stored in non-volatile memory when it successfully completes.
Other approaches considering non-volatile processors, that not only integrate non-volatile memories but also non-volatile registers and flip-flops, have been proposed \cite{Ma2015}. However, the cost in terms of hardware (increased power consumption, increased area, and decreased frequency) results in significant software slowdowns and complexity overhead.

Alpaca \cite{Maeng2017}, Mayfly \cite{Hester2017c} and InK \cite{Yildirim2018} are the most relevant state-of-the-art task-based computing models and schedulers. The first two only consider static task flows, and if any task cannot be completed due to the energy level at that time, it will be executed again. 
However, if a specific task is not able to be completed or if the energy conditions of the capacitor or energy harvester change, any other tasks will starve, waiting for the current one to be tried over and over again. To overcome this problem, InK \cite{Yildirim2018} considers a dynamic scheduler based on priorities and event-triggers (e.g., timers, energy level triggers, sensor value triggers), which are defined in advance by the programmer. This allows the application to adapt to changes in available energy and variations in sensing behavior. 
However, it places the entire burden of adapting the application logic and task selection on the programmer. This requires them to have in-depth knowledge on the energy consumption of tasks, as well as the energy life-cycle of the device, which is generally not known before deployment of the device. If the device does not know how much energy will be available in the future, it can spend energy on a task or chain of tasks without knowing if it will have enough energy for completing it before the deadline. If the harvested energy is insufficient, tasks will not be completed, and energy and time will be wasted. To overcome this issue, and fully automate the problem of task selection, in this paper we propose an energy-aware task scheduler. We consider the task chains’ priorities and deadlines, but also the available energy of the device, the energy cost of the tasks and the (predicted) energy harvesting budget, to decide which task should be completed first, making it more intelligent and resource-efficient. Although Mayfly \cite{Hester2017b} and InK \cite{Yildirim2018} consider deadlines and data freshness in the form of expiration timers, they do not consider deadlines across the different tasks of a chain.
AsTAR \cite{Yang2019} also presents an energy-aware task scheduler and an associated reference platform that aims to lower the burden of developing sustainable applications through self-adaptative task scheduling. It does not need any pre-configuration and supports platform heterogeneity. However, it does not consider different types of application requirements nor different priorities in their design. More recently, Islam \textit{et al.} \cite{Islam2020} have proposed two scheduling algorithms for batteryless devices where the energy of the capacitor and the deadline of the tasks are taken into account. However, they assume that the capacitor follows a linear charging behaviour, harvesting and computing are exclusive and applications have all the same priority. In our work, we follow a different approach, where we use an exponential capacitor charging model, and we let the harvester and the device to work simultaneously. We also define different task priorities, to allow more power hungry tasks to be executed if they are more beneficial for the specific use case.
 
 Other scheduling algorithms have also been proposed. Caruso \textit{et al.} proposed a dynamic programming algorithm for the optimization of the scheduling of the tasks in IoT devices that harvest energy by means of a solar panel \cite{Caruso2018}. They used estimations of the solar energy that is produced in each slot of time to compute the optimal scheduling in advance. 
 \textcolor{black}{SolarCore \cite{Li2011} also presents a solution for solar energy harvesting. It includes a power management scheme to optimize the power obtained from a solar panel thanks to the maximal power provisioning control and workload optimization.}
 However, in our work we present a more generic task scheduling solution that does not rely on any specific energy source \textcolor{black}{and whose main constraint is the energy scarcity available in the capacitor.}
 A simple scheduler model that does not consider different tasks, nor a harvesting source is proposed in \cite{Miguel2018}, where authors explain the importance of taking into account the energy consumption of the memory for backing up and restoring the data to and from non-volatile memory when a power failure occurs. They claim that expending energy on instructions whose output is not saved before a power outage is wasteful. The authors in \cite{Yang2016} present the modified earliest deadline first (MEDF) algorithm based on super capacitors and energy harvesting that takes into account energy  and time constraints of the tasks. However, in their algorithm they do not avoid energy violations (when the voltage of the capacitor falls below the turn-off threshold), but only count how many occur. Counting how many energy violations happen without avoiding them or considering the device needs to turn-off is not realistic and something we address in this paper. Furthermore, they assume the tasks to be independent and nonpreemptable, while we not only avoid power failures but also consider tasks dependency in our work.

A more practical scheduling algorithm has been presented in \cite{Escolar2018}, where the authors optimally set the overall node power consumption based on the utility, and on the energy required by tasks. It is implemented on an Arduino node, equipped with a small (portable) solar panel, and attached to a small wind turbine. In \cite{Srbinovski2016}, Srbinovski \textit{et al.} present an energy aware adaptative sampling algorithm, where the node manages its activity in the network according to its energy levels. However, the user needs to define the critical battery level at which it becomes energy conservative by reducing the sampling rate. This is the same approach followed by AsTAR \cite{Yang2019}. 
\textcolor{black}{In a more general scenario where a server farm needs to be optimized, Blink \cite{Sharma2011} proposes to frequently adjust the servers' duty cycle (to turn on and off the servers) to adapt to power variations, while maintaining a certain synchronization between them when needed.} \textcolor{black}{And in \cite{Li2019} an energy-aware scheduling algorithm that is able to configure the hardware of a Field-Programmable Gate Array (FPGA) based on the weather forecast solar energy available is proposed. However, we look at the optimal scheduling of application tasks on a constrained IoT device without batteries, which is a problem with significantly different constraints and requirements.}

Although ILP-based approaches have already been proposed, they are not intended to work for batteryless devices. In fact, \cite{Baruah2008} presented an ILP approach to be used with multiprocessor partitioned scheduling and assume that any task may be interrupted at any instant in time, and its execution resumed later with no cost or penalty, which might not be realistic in batteryless devices. 
The same assumption is also used in \cite{Moser2007}, where the authors present an optimal scheduling algorithm for rechargeable batteries or supercapacitors, although their energy model is very simplistic.
\textcolor{black}{In \cite{Lee2019} a task scheduling algorithm for Simultaneous Wireless Information and Power Transfer (SWIPT) IoT devices is proposed, where only one device can be charged at a time. As we target environmental sources (e.g., solar), all devices could be charged at the same time. Although in their evaluation they use a rechargeable battery and a capacitor as storage elements, they consider that both follow a linear charging behaviour.   }
We have also presented an energy-aware algorithm for batteryless LoRaWAN devices using energy harvesting, where we evaluate the performance of these constrained devices when allowing sleeping between tasks or letting them turn off \cite{Sabovic2020}. We showed that sleeping between tasks normally performs better, and for this reason, in this work we present a more generic task scheduler where we follow this approach. 
Finally, in order to reduce energy consumption, task offloading could be used, as authors in \cite{Bi2020} have used in smart mobile devices. The task offloading could be combined with our scheduling algorithm for constrained batteryless IoT devices where, according to the available energy, the scheduler should decide whether to transmit data towards the edge cloud where computing should take place rather than performing computations on the device itself.

 \section{System model}\label{sec:model}
In this section we give a brief overview of the considered batteryless IoT device model (more details are provided in our previous work \cite{Delgado2020})
and how the energy-aware task scheduler for these batteryless devices works.

\subsection{Batteryless model}\label{sec:batterylessmodel}
\begin{figure}[t]
\centerline{\includegraphics[width=0.75\columnwidth]{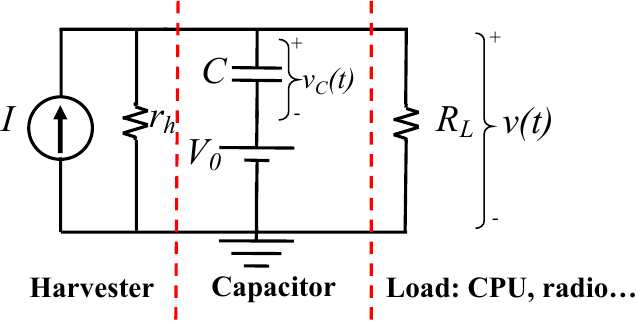}}
\caption{Electrical circuit model of a batteryless IoT device using a current source energy harvester}
\label{fig:Circuit}
\end{figure}

Batteryless IoT devices are equipped with a harvester mechanism, a capacitor, an MCU, a radio unit and the needed peripherals. In order to model the behaviour of these devices, we have considered the electrical circuit presented in \cite{Delgado2020} and shown in Figure~\ref{fig:Circuit}, where the circuit is
divided into three main parts: the harvester (source of the energy), the capacitor (storage of the energy) and the load (consumer of the energy: MCU, radio, peripherals).

We have considered a generic and simplified approach, where only the generated power is taken into account. Since photovoltaic cells are modeled as current harvester sources \cite{Mahmoud2012}, we have considered the current source model presented in \cite{Delgado2020}, where the harvester is modeled as an ideal current source composed by a real current source in parallel with an internal resistance (denoted by $I$ and $r_h$, respectively).
The parallel resistance $r_h$ (in $\Omega$) limits the power of the harvester and its value is calculated using the following equation: 
\begin{equation}
 r_h = \frac{V_{max}^2}{P_{harvester}}
\label{eq:ri} 
\end{equation} 

\noindent where $P_{harvester}$ is the power of the harvester source, which can vary greatly depending on the type of energy harvesting considered (e.g., up to $1 mW/cm^2$ for indoor natural light, and up to $100 mW/cm^2$ for outdoor sun) \cite{Shirvanimoghaddam2018}, and $V_{max}$ (in Volts) is the maximum voltage supported by the circuit elements, which is determined by the load. And the value of the current I (in Amperes) can be calculated as follows:
\begin{equation}
I = \frac{V_{max}}{r_h}
\end{equation} \label{eq:I}

The capacitor is the part of the circuit where the energy is stored. As shown in Figure~\ref{fig:behaviour}, the behaviour of the system is a succession of intervals, where the capacitor is being charged or discharged. Each interval is characterized by a specific state of the load components (e.g., MCU is active and radio is transmitting). 
We characterize the voltage of the capacitor throughout each interval using $V_0$ and $v_C(t)$. $V_0$ represents the initial voltage of the capacitor at the beginning of the interval (i.e., time $t_0$), and $v_C(t)$ is the temporal evolution of said voltage at time $t$ (relative to $t_0$). Both $V_0$ and $v_C(t)$ are included in the circuit as an ideal voltage source and the voltage over time of an ideal capacitor respectively, as shown in Figure~\ref{fig:Circuit}.

The load of the model corresponds to the set of components that consume the stored energy in the capacitor per task being executed, such as 
the MCU, radio or sensors. Each of these components is characterized by a specific power consumption in each of its states (e.g., active, sleeping, off).
Therefore, they can be modeled as a load resistance denoted by $R_L$ (in $\Omega$), which can be calculated as follows:
\begin{equation}
 R_L = \frac{E}{I_{load}}
 \label{eq:RL} 
\end{equation} 

\noindent where $E$ and $I_{load}$ can be defined either theoretically or empirically. Theoretically, $E$ is the operating voltage of the circuit elements, which is given in the datasheet, and $I_{load}$ is the sum current consumption of all components for the specific task to be executed, which can also be found in the corresponding datasheets. In order to calculate it empirically, $E$ is the corresponding voltage value at which the device is being powered while $I_{load}$ is the current consumption measured per each task. 
$R_L$ thus varies across different tasks depending on the state of each component (i.e., radio, MCU, or sensors) during that specific task. Please note that $I_{load}$ is renamed as $e_j$ to be consistent with the scheduler formulation in the energy-aware task scheduler in Section \ref{sec:optimization}.

To determine if the device has enough energy at a specific time $t$ to perform its tasks (e.g., transmit data), it is needed to calculate the voltage across the load of the model $v(t)$:

\begin{equation}
 v(t) = I R_{eq}(1-e^{(\frac{-t}{R_{eq}C})})+V_0e^{(\frac{-t}{R_{eq}C})}
 \label{eq:V_Current} 
\end{equation} 

\noindent where $C$ is the capacitance in Farads, $t$ is the time (in seconds) spent in the current task, and $R_{eq}$ is the equivalent resistance of the circuit (in $\Omega$), computed as:
\begin{equation}
 R_{eq}=\frac{R_L r_h}{R_L + r_h}
 \label{eq:Req} 
\end{equation} 

The value of $v(t)$ provides the voltage available in the load, which will be used to determine if a specific task (e.g., transmit, listen, sense) can be performed during an interval, according to the needed time $t$ it will take, the energy harvesting rate $P_{harvester}$, the specific load $I_{load}$, and the capacitor voltage $V_0$ at the start of the execution of such tasks. Note that $v(t)$ can be increasing or decreasing depending of the specific parameters, and if it goes below the turn-off threshold, the device (which is represented as the load in Figure~\ref{fig:Circuit}) will turn off.

\subsection{Energy-aware task scheduling concept}\label{sec:EATS}
As mention in Section \ref{sec:sota}, in this work we propose a new optimization framework that uses two approaches: we first divide tasks into atomic subtasks and second, we use duty cycling, meaning that we let the device sleep to replenish energy when there are no tasks to be executed or if there is not enough energy, so it will harvest more energy benefiting the successful completion of future tasks.

\begin{figure}[t]
\centerline{\includegraphics[width=1\columnwidth]{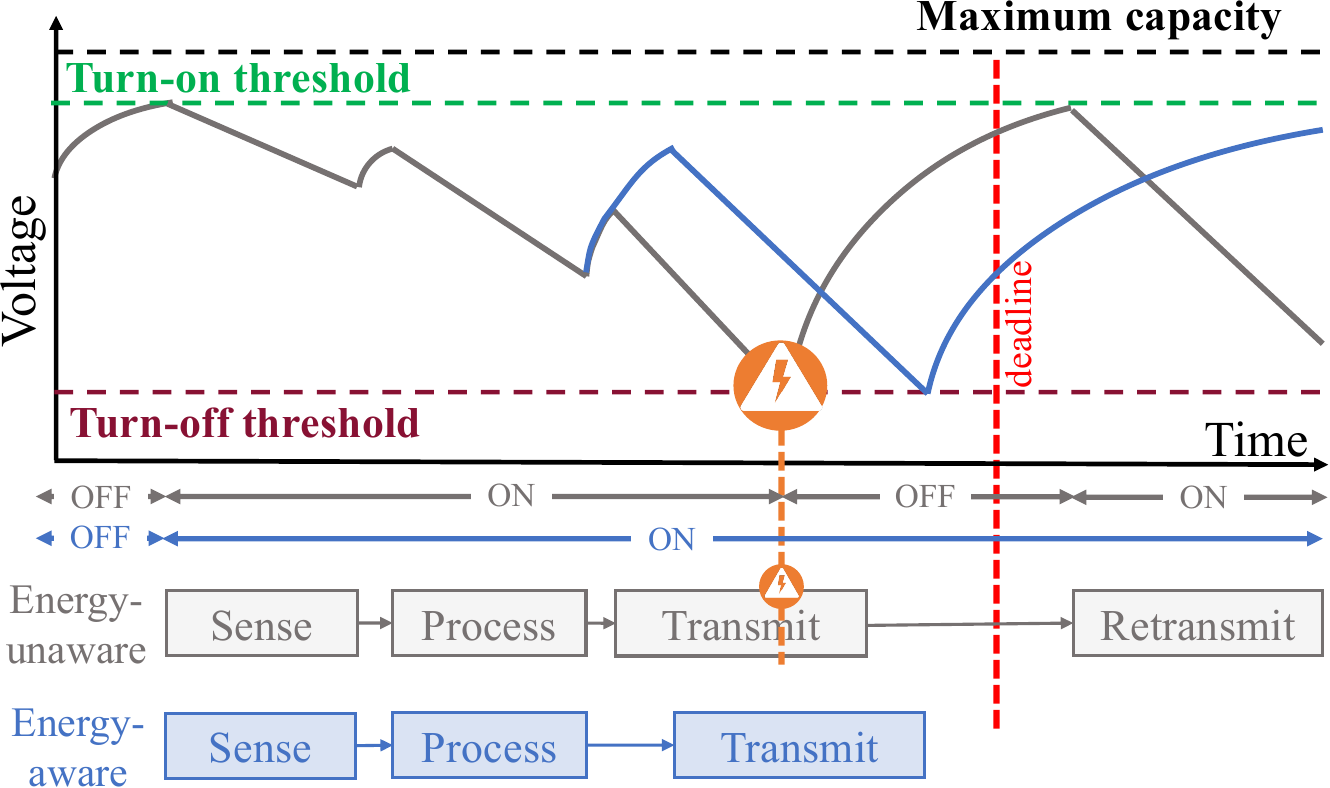}}
\caption{Energy-unaware vs energy-aware task schedulers in batteryless devices}
\label{fig:EAScheduler}
\end{figure}

The state-of-the-art schedulers introduced in Section \ref{sec:sota}, often do not consider the energy in their algorithms, while we argue in batteryless devices, energy is the main concern. In Figure \ref{fig:EAScheduler} we show the comparison of the behavior of the state-of-the-art approach and our energy-aware approach. In this example, the temperature update application is composed of three tasks: sense, process and transmit. As can be seen, it first performs the sensing task, which corresponds to take measurements from the environment, for example, reading the temperature. Then, the data is processed and transmitted to the central controller in the following two tasks, completing the task chain for this event. However, existing task schedulers for intermittent computing select the next task to execute based on simple priority rules \cite{Yildirim2018}, leading to potential deadlocks or wasting of the scarce available energy on task chains that cannot be completed on time anyway. 

As also reflected in Figure \ref{fig:EAScheduler}, we are targeting applications and tasks with deadlines, since for example, data freshness is important in monitoring applications (e.g., sending a heart rate measurement of several minutes old might be useless). So we consider a device trying to complete a task chain within a deadline. If it is energy-unaware (depicted in grey), it will try to complete every task one after another. However, when it tries to transmit the data in the example, there is not enough energy to complete this task, and a power failure occurs. In this case, the device needs to harvest additional energy before it reaches its voltage turn-on threshold and is able to retransmit the data. However, by the time this happens, the deadline is missed (depicted by the red line), which means that the data is not \textit{"fresh"} anymore, and useless to the backend application that uses~it. 

To solve this problem, our energy-aware scheduler instead decides to wait till enough energy is available to be able to transmit the data before the deadline. In order to do so, in this work we are assuming we have knowledge of the energy consumption of the tasks and the energy that is being harvested over a certain look-ahead time window. The first assumption is easy to fulfill since we can either use the data sheet consumption values of the reference platforms used or by measuring it before device deployment. Moreover, we can assume we have the knowledge of the future energy that can be harvested for some predictable and controllable sources such as RF or indoor light \cite{Shaikh2016}.

The proposed energy-aware task scheduler will determine which task should be executed and when, according not only to the three main batteryless components: the energy available in the capacitor, the energy that is being harvested, and the the energy consumption of the device performing the tasks; but also the tasks requirements (arrival time, execution time, priority, deadline and order of the tasks). The main goal of the energy-aware scheduler is to execute the maximum number of tasks, prioritizing its priority status. The most common types of tasks on the IoT node include sampling of information, processing the data, data transmission, data reception or the use of the actuators.

We consider a task as a sequence of atomic operations that are executed on a node. Typical IoT applications are for example report sensor values, relay data or use an actuator. We divide these applications into tasks. For example, report sensor values can be divided into sense the environmental variable (e.g., temperature) and transmit the data, relay data can be divided into receive and transmit and use an actuator can be divided into receive the order and actuate. Every tasks will be characterized by its arrival time, execution time, deadline, priority, order and energy consumption. We consider that the energy consumption of the tasks already take into account the consumption of the memory for back up and restore, as mentioned in \cite{Miguel2018}. It is also important to mention that although some tasks such as receiving are very technology specific (LoRaWAN and BLE have different behaviours), they can also be considered as a single task where the different parameters of the optimization framework (i.e., execution time will determine the whole BLE transmission over the three channels) will need to be defined accordingly.

\begin{figure}[t]
\centerline{\includegraphics[width=0.56\columnwidth]{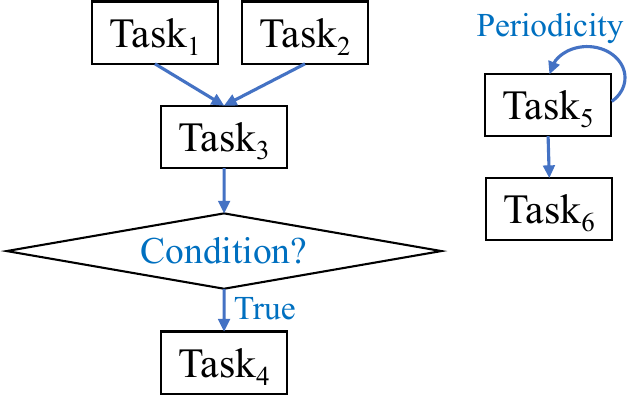}}
\caption{General overview of the atomic tasks model}
\label{fig:GenTasks}
\end{figure}

In Figure~\ref{fig:GenTasks} we show an example of how we have modeled the atomic tasks. Every task is defined by execution time, deadline, priority and energy consumption. Some of the tasks are also characterized by an order. This is the case for $Task_3$, that has two parents which are $Task_1$ and $Task_2$, or $Task_6$, which has a $Task_5$ as a parent. Finally, the arrival time will depend on the specific task. For example, the arrival time of $Task_1$ and $Task_2$ should be given as an input, while the arrival time of $Task_5$ is periodic. Furthermore, $Task_4$ will only arrive if the condition of $Task_3$ is satisfied.

\section{Optimization framework}\label{sec:optimization}
This section formally describes the energy-aware task scheduling problem which aims to maximize the successful task execution rate (weighted by task priority). We formulate the optimization problem as a MILP that uses, among others, the energy that its being harvested, the tasks execution time and the tasks energy consumption as input and decides on the optimal way of executing the tasks avoiding power failures and missing deadlines. 
The MILP will provide the globally optimal if we assume a total knowledge of the expected tasks to be executed and the energy that is being harvested. However, 
given knowledge about the tasks to execute and (predicted) energy harvested over a certain look-ahead time window, the scheduler can calculate a sub-optimal solution that is optimal over that specific time window.
Furthermore, since energy harvesting is normally not known in advance, energy harvesting predictors can be used \cite{Adu-Manu2018}.
The remainder of this section describes the different aspects of the MILP formulation and Table \ref{tab:optParam} summarizes the notations used.

\begin{table}[t]
\caption{Set of parameters of the optimization framework}
\label{tab:optParam}
\centering
\begin{tabular}{ll}
\hline
\textbf{Parameter}	& \textbf{Definition}	\\
\hline
$T = \{t_1, t_2, ..., t_{|T|}\}$		& Set of time instants; subscript index $t$ \\
                    & refers to time instant $t_t$\\
$A = \{a_1, a_2, ..., a_{|A|}\}$		& Set of tasks to be executed; subscript \\
                    & index $j$ refers to task $a_j$\\
$R_j = \{p_j, te_j, ta_j, d_j , e_j\}$ & Requirement vector of task $j$ (priority, \\
                    & execution time, arrival time, deadline \\
& and current consumption)\\
$P_{j} $& Subset of tasks that need to be  \\
                    & executed before task $j$ is executed\\
$y_{jt}$		& Binary decision variable indicating \\
                    & if task $j$ starts executing at time $t$\\
$z_{jt}$		& Binary decision variable indicating \\
                    & if task $j$ is being executed at time $t$\\
$V_{t}$		& The voltage of the capacitor at time \\
                    & instant $t$\\
\hline
\end{tabular}
\end{table}

\subsection{Input variables}
Let $T = \{t_1, t_2, ..., t_{|T|}\}$ be the ordered set of time instants where the set $A = \{a_1, a_2, ..., a_{|A|}\}$ of tasks need to be scheduled. To simplify notation, in the following we will use the subscript index $t$ to refer to a instant of time $t_t$ and the subscript index $j$ to refer to a task $a_j$. Each task $j \in A$ is characterized by its requirement vector $R_j = \{p_j, te_j, ta_j, d_j, e_j\}$, where  $p_{j}$ is the priority, 
$te_{j}$ is the execution time of the task,
$ta_j$ is the arrival time, meaning that the task cannot be scheduled before this time, and 
$d_{j}$ is the task deadline to guarantee the freshness and usability of the data and tasks outputs, and 
$e_{j}$ is the average current consumption of the task (which is considered constant per each task).
Furthermore, in order to guarantee the task chain order, every task $j \in A$ has a set of parents $P_{j} \subset A$, which contains the set of tasks that need to be executed before the execution of task $j$. The set of parents $P_{j}$ can also be empty.

\subsection{Decision variables} \label{decisionVbles}
There are two decision variables in the MILP, $y_{jt}$ and $z_{jt} \quad \forall j \in A, \forall t \in T$, which represent the specific task scheduler decisions. While $y_{jt}$ is the binary variable indicating if task $j$ starts executing at time $t$, $z_{jt}$ is the binary variable indicating if task $j$ is being executed at time $t$. This means that if a task $j$ starts executing at time $t$ and its execution time is $te_j = 2$, $y_{jt} = 1$, $z_{jt} = 1$, $z_{jt+1} = 1$ and for the rest of the elements in the set $T$, $y_{jt}$ and $y_{jt}$ are equal to 0.

\subsection{Objective function and constraints}

The main goal of the energy-aware task scheduler is to maximize the number of tasks successfully scheduled multiplied by their priority:
\begin{equation}
	\label{eq:maximiza}
	\max \sum_{j \in A} \sum_{t \in T} y_{jt} \times p_{j} 
\end{equation}

The presented objective function is restricted by some constraints. First of all, two tasks cannot be scheduled at the same time, but also a task $j$ should only be scheduled once, as specified in Equations \ref{eq:const2} and \ref{eq:const3} respectively:
\begin{equation}
	\label{eq:const2}
	\sum_{j \in A}  z_{jt} \leq 1  \quad \forall t \in T  
\end{equation}

\begin{equation}
	\label{eq:const3}
	\sum_{t \in T}  y_{jt} \leq 1  \quad \forall j \in A  
\end{equation}

We need to ensure that task $j$ is only successful if it is deployed for all its execution time ($te_j$) (see Equation \ref{eq:const4}) but also that if a task $j$ starts its execution at time $t$, it is being executed from time $t$ until $t+te_{j}$, as Equation \ref{eq:const5} defines.

\begin{equation}
	\label{eq:const4}
	\sum_{t \in T} y_{jt} = \frac{1}{te_{j}} \sum_{t \in T} z_{jt} \quad \forall j \in A  
\end{equation}

\begin{equation}
	\label{eq:const5}
	z_{jt} = \sum_{u \in [t - te_{j} + 1, t]} y_{ju} \quad \forall j \in A, \forall t \in T 
\end{equation}

Furthermore, task $j$ cannot be scheduled before its arrival nor after its deadline, and Equation \ref{eq:const6} guarantees it.

\begin{equation}
	\label{eq:const6}
\sum_{ d_{j} < t < ta_{j}} z_{jt} = 0 \quad \forall j \in A
\end{equation}

Every task $j$ needs to be executed after all its parents in the set $P_{j}$ have finished their executions:

\begin{equation}
	\label{eq:const7}
\sum_{u \leq t} y_{ju} \leq \sum_{u \leq t} y_{pu} \quad \forall j \in A , \forall p \in P_{j}, \forall t \in T
\end{equation}

As mentioned in Section \ref{sec:batterylessmodel}, we are assuming every device is equipped with a harvester and a capacitor to store the harvested energy from the environment. The voltage across the capacitor for every instant of time $t$ is defined by the continuous variable $V_t \quad \forall t \in T$. It is important to note that $V_t$ is the discrete time version of $v(t)$ of Equation \ref{eq:V_Current}.
We need to guarantee that the voltage across the capacitor is enough to execute the scheduled tasks. For this reason, we first need to ensure that this voltage remains between the minimum and maximum supported values ($V_{min}$ and $V_{max}$). $V_{max}$ was already defined in Section \ref{sec:batterylessmodel}, and $V_{min}$ corresponds to the voltage turn-off threshold of Figure \ref{fig:behaviour} and Figure \ref{fig:EAScheduler}.

\begin{equation}
	\label{eq:constVoltage}
	V_{min} \leq V_t \leq V_{max}  \quad \forall t \in T  
\end{equation}
	
The voltage across the capacitor needs to be calculated for every time instant and depends on the specific energy harvested and the energy consumed by the tasks per every time instant. Applying Equation \ref{eq:V_Current} to our variables, we obtain the following constraint:

\begin{gather}
\label{eq:constV1} 
V_t =  \sum_{j \in A} z_{jt} \times  \\
\left(I_t R_{eq_{j,t}}(1-e^{(\frac{-\bigtriangleup t}{R_{eq_{j,t}}C})})
+V_{t-1}e^{(\frac{-\bigtriangleup t}{R_{eq_{j,t}}C})} \right)     \quad \forall t \in T
\notag
\end{gather}

\noindent where $\bigtriangleup t$ corresponds to the amount of time (in seconds) between two instants of time in $T$ (i.e., $\bigtriangleup t = time(t_t - t_{t-1})$), $V_{-1}$ is the known initial voltage of the capacitor, $C$ is the capacitance in Farads of the capacitor. As explained in Section \ref{sec:batterylessmodel}, the harvester is modeled as a real current source composed of an ideal current source and a parallel resistance. Since these values can vary over time, we now denoted them by $I_t$ and $r_{h_t}$, respectively. The value of $I_t$ (in Amperes) is calculated as follows:

\begin{equation}
\label{eq:It} 
I_t = \frac{V_{max}}{r_{h_t} } \quad \forall t \in T
\end{equation} 

\noindent where the parallel resistance $r_{h_t}$ (in $\Omega$), that limits the power of the harvester, is calculated using the following equation (similar to Equation \ref{eq:ri}): 
\begin{equation}
 r_{h_t} = \frac{V_{max}^2}{P_{harvester,t}} \quad \forall t \in T
\end{equation} \label{eq:rht} 
where $P_{harvester,t}$ is the power of the harvester source at time instant t, in Watts. Finally, $R_{eq_{j,t}}$ is the equivalent resistance of the circuit (in $\Omega$) at time instant t, which depends on the specific task $j$ that is scheduled at that time, and is computed as:
\begin{equation}
R_{eq_{j,t}}=\frac{R_{L_j}  r_{h_t}}{R_{L_j} + r_{h_t}} \quad \forall j \in A, \forall t \in T
 \label{eq:Reqjt} 
\end{equation} 
where $R_{L_{j}}$ (in $\Omega$) is the load of the model (it corresponds to the set of components that consume the stored energy in the capacitor when executing the scheduled task $j$), which can be calculated as follows:
\begin{equation}
 R_{L_j} = \frac{E}{e_{j}} \quad \forall j \in A
 \label{eq:RLj} 
\end{equation} 
where $E$ is the operating voltage of the circuit elements and $e_{j}$ is the current consumption of task $j$, defined in the $R_j$ vector.

In order to compute Equation \ref{eq:constV1}, it is needed to define all $V_t$ for all values of $t$. However, if at a certain point in time the device is sleeping and no task is being executed, $z_{jt}=0$ for all values of $j$. For this reason, we need to adapt and reformulate Equation~\ref{eq:constV1} as follows:

\begin{gather}
V_t =  \notag \\
\sum_{j \in A} z_{jt} \times \left( I_t R_{eq_{j,t}}(1-e^{(\frac{-\bigtriangleup t}{R_{eq_{j,t}}C})})+ V_{t-1}e^{(\frac{-\bigtriangleup t}{R_{eq_{j,t}}C})}\right)   + \notag \\
(1 - \sum_{j \in A}   z_{jt} ) \times \left(  I_t R_{eq_{s,t}}(1-e^{(\frac{-\bigtriangleup t}{R_{eq_{s,t}}C})})+V_{t-1}e^{(\frac{-\bigtriangleup t}{R_{eq_{s,t}}C})}\right) \notag \\
\quad \forall t \in T
	\label{eq:constV2}
\end{gather}

where $R_{eq_{s,t}}$ is calculated when the device is in sleep mode, which means we use the current consumption of the sleep mode for calculating it. And we can reformulate Equation \ref{eq:constV2} as follows:

\begin{gather}
V_t - \sum_{j \in A}  z_{jt} \times \notag \\
 \left( I_t R_{eq_{j,t}}(1-e^{(\frac{-\bigtriangleup t}{R_{eq_{j,t}}C})}) -  I_t R_{eq_{s,t}}(1-e^{(\frac{-\bigtriangleup t}{R_{eq_{s,t}}C})}) \right) - 
\notag \\ \sum_{j \in A} z_{jt} \times V_{t-1} \times \left(e^{(\frac{-\bigtriangleup t}{R_{eq_{j,t}}C})} - e^{(\frac{-\bigtriangleup t}{R_{eq_{s,t}}C})}  \right) - \notag \\
V_{t-1}e^{(\frac{-\bigtriangleup t}{R_{eq_{s,t}}C})} - 
I_t R_{eq_{s,t}}(1-e^{(\frac{-\bigtriangleup t}{R_{eq_{s,t}}C})}) = 0 \quad \forall t \in T
\label{eq:constV3}
\end{gather}

Considering that $V_{t-1}$ is dependent on the decision variable $z_{jt}$, the multiplication $z_{jt} \times V_{t-1}$ is no longer linear. To linearize it, we can define a new continuous variable $\varUpsilon_{jt} = z_{jt} \times V_{t-1}$, and Equation \ref{eq:constV3} can be reformulated as:

\begin{gather}
V_t - \sum_{j \in A} z_{jt} \times  \notag \\
\left( I_t R_{eq_{j,t}}(1-e^{(\frac{-\bigtriangleup t}{R_{eq_{j,t}}C})}) -  I_t R_{eq_{s,t}}(1-e^{(\frac{-\bigtriangleup t}{R_{eq_{s,t}}C})}) \right) - \notag \\
\sum_{j \in A}  \varUpsilon_{jt} \times \left(e^{(\frac{-\bigtriangleup t}{R_{eq_{j,t}}C})} - e^{(\frac{-\bigtriangleup t}{R_{eq_{s,t}}C})}  \right) - V_{t-1}e^{(\frac{-\bigtriangleup t}{R_{eq_{s,t}}C})} = \notag \\
  I_t R_{eq_{s,t}}(1-e^{(\frac{-\bigtriangleup t}{R_{eq_{s,t}}C})})  \quad \forall t \in T
\label{eq:constV4}
\end{gather}

\textcolor{black}{where, since $V_{t-1}$ is bounded below by \textit{zero} and above by $V_{max}$, the variable $\varUpsilon_{jt}$ needs to fulfill the following restrictions:}

\begin{equation}
	\label{eq:const16}
	\varUpsilon_{jt} \leq z_{jt} \times V_{max} \quad \forall j \in A, \forall t \in T
\end{equation}

\begin{equation}
	\label{eq:const17}
	\varUpsilon_{jt} \leq V_{t-1} \quad \forall j \in A , \forall t \in T
\end{equation}

\begin{equation}
	\label{eq:const18}
	\varUpsilon_{jt} \geq V_{t-1} - (1 - z_{jt}) \times V_{max} \quad \forall j \in A , \forall t \in T
\end{equation}

\begin{equation}
	\label{eq:const19}
	\varUpsilon_{jt} \geq 0 \quad \forall j \in A , \forall t \in T
\end{equation}

To sum up, we define our MILP scheduler by the objective function in Equation \ref{eq:maximiza} subject to the constraints of the Equations \ref{eq:const2}, \ref{eq:const3}, \ref{eq:const4}, \ref{eq:const5}, \ref{eq:const6}, \ref{eq:const7}, \ref{eq:constVoltage}, \ref{eq:constV4}, \ref{eq:const16}, \ref{eq:const17}, \ref{eq:const18} and \ref{eq:const19}, and where Equations \ref{eq:It} - \ref{eq:RLj} are not constraints but helpers for Equation \ref{eq:constV4}.

\section{Evaluation}\label{sec:evaluation}
 
In this section we evaluate the performance of the proposed energy-aware task scheduler. We first introduce the simulation setup and the methodology used to evaluate the optimal algorithm. Then, in order to validate its behaviour, we compare it against one of state-of-the-art solution. Finally, we provide some insights on how long in the future the behaviour of the batteryless device needs to be predicted
in order to get the best performance and avoid power failures.

\subsection{Simulation setup and methodology}\label{sec:methodology}

As detailed in Section \ref{sec:batterylessmodel}, we have considered a batteryless device, composed of a harvester (e.g., a photovoltaic cell) that is able to continuously harvest a power of $P_{harvester}$  \cite{Delgado2020}, and that can store it when not used in its capacitor, with a capacitance $C$. Let us name the harvested power as~$PH$. 

\begin{figure}[t]
\centerline{\includegraphics[width=0.95\columnwidth]{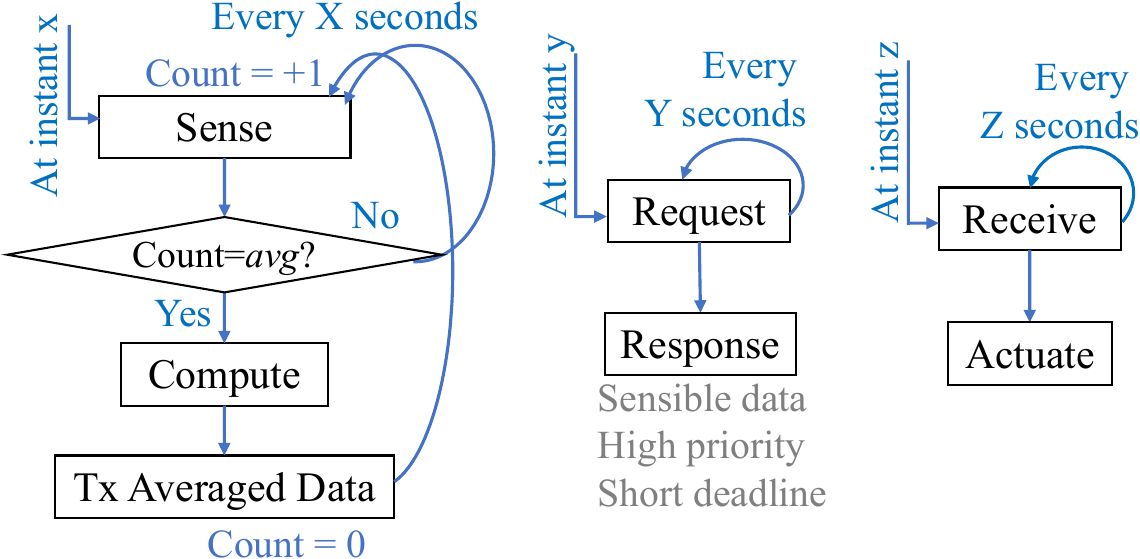}}
\caption{Detailed tasks of the Smart Building Application}
\label{fig:SBA}
\end{figure}

We consider a smart building application, composed of three main commands: sense and transmit the average data, request-response and receive$\&$actuate, as can be seen in Figure \ref{fig:SBA}. The first command is able to sense an environmental variable (i.e., temperature) periodically every $X$ seconds, compute the average of $avg$ samples and transmit it. Secondly, if the device receives a request, it will immediately answer to it with the response task. This task needs to be executed within a short deadline as it is considered containing sensible data. Finally, we have also considered that the device can receive a message to enable one of its peripherals (such as an alarming LED). In the detailed diagram it is possible to see what are the periodicities considered and the parents of the atomic tasks. For example, the Request task will have a $Y$ seconds periodicity and the Response task has only one parent which is Request. 

In order to be able to communicate, the device is also equipped with a Bluetooth Low Energy (BLE) radio. It is important to note that in order to simplify the problem, we have considered that a transmission in the three advertisement channels of BLE is an atomic task, and therefore we have considered the average energy consumption of the complete transmission, as no differences in the results were observed after making this simplification. Furthermore, every considered task also takes the energy consumption of the backup memory and of restoring it from memory into account, as already mentioned in Section \ref{sec:EATS}.

\begin{table*}[t]
\caption{Smart Building Tasks considered and their parameters}
\label{tab:SBAparameters}
\centering
\begin{tabular}{cccccc}
\hline
\textbf{Task, $j$} & \textbf{Priority, $p_j$}	& \textbf{Execution Time, $te_j$}& \textbf{Current Consumption, $e_j$}& \textbf{Deadline, $d_j$}&  \textbf{Comments}\\
\hline
Sense&      1			& 0.03s & 1.7mA & $1/3X$ &Periodically sense\\
Compute&	3			& 0.01s & 1mA & 1s &Average the sensed data\\
Tx&	3			& 0.19s & 4.36mA & 1s &Tx the averaged data  (4dBm)\\
Request &	8			& 0.21s & 4.61mA & 0.2s &Receive a request\\
Response &	10			& 0.19s & 4.36mA &0.02s& Send a response  (4dBm)\\
Receive &	8			& 0.21s & 4.61mA & 0.2s & Receive an order\\
Actuate &	8			& 0.05s & 9mA & 1s& Use an actuator\\
\hline
\end{tabular}
\end{table*}

For our simulations, we have considered that the sensing task is executed every second ($X=1s$) and its first arrival is at time 0 ($x=0s$). Additionally, we would like to compute the average samples every 5 samples ($avg=5$). The two other commands of the diagram of Figure \ref{fig:SBA}, Request and Receive, arrive at the instants 1 and 3 seconds, and their periodicity is 2 and 5 seconds, respectively. This is $y=1s$, $z=3s$, $Y=3$ and $Z=5$. 

In Table \ref{tab:SBAparameters} we show all the parameters considered for the atomic tasks. The priority has been determined based on the type of the application, sensing periodically is not as critical as responding to an urgent Request. The values of the current consumption and execution time are based on the specifications of the Nordic nRF52840 \cite{mcu}, and $e_j$ is the theoretical value for obtaining $I_{load}$ (see Section \ref{sec:batterylessmodel}). As such, $V_{min}$ has been defined as 1.8V (minimum operating voltage) and $E$ and $V_{max}$ have been defined as 3.3V (typical operating voltage). The value of the deadline has been taken according to the urgency of the tasks. For example, executing a Response for the Request is considered critical, and therefore its deadline is short. In contrast, the sensing task has a deadline which depends on its periodicity, which means that if its not being sensed in the first third of its periodicity, that data is not longer \textit{"fresh"}. The needed parameters of the batteryless devices are shown in Table \ref{tab:setup}, unless explicitly specified.

\begin{table}[t]
\caption{Experiment setup}
\label{tab:setup}
\centering
\begin{tabular}{cc}
\hline
\textbf{Definition}	& \textbf{Value}	\\
\hline
$V_{min}$		& 1.8V	 \\
$V_{max}$		& 3.3V		 \\
$V_{-1}$ (initial capacitor voltage)		& 2.2V		 \\
$C$	& 		4.7mF	 \\
$P_{harvester,t}$	& Constant $\forall t \in T$			 \\
$\bigtriangleup t$	& 0.01s			 \\
Sleep Current Consumption	& 0.1mA			 \\
Turn On Current Consumption & 3mA \\
Turn On Time & 0.1s \\
\hline
\end{tabular}
\end{table}

To solve the optimization problem described in Section~\ref{sec:optimization}, we have used Gurobi Optimization\footnote{https://www.gurobi.com/}. 
The output of the optimizer is then fed into the event-based simulator in which we evaluate the proposed solution. This event-based simulator has been implemented in C++, and simulates the energy level of the batteryless device according to the expected energy to be harvested, and the scheduled tasks.  
Finally, and as mentioned before, for simplicity of the analysis, we assume a constant energy harvesting rate during a single experiment, which is in line with the output of a buck regulator \cite{Delgado2020}.
Even if the input of the buck regulator is not constant, it provides a constant output. However, both the optimizer and the simulator allow to work with time-varying harvesting power.

\subsection{Energy-aware scheduler validation}
In order to validate our energy-aware scheduler (E-Aware), we compare its behavior against InK~\cite{Yildirim2018}, one of the most complete schedulers for batteryless devices of the state-of-the-art. InK is a dynamic scheduler based on priorities and deadlines, however, it is not energy-aware. Although AsTAR \cite{Yang2019} is energy-aware, it is only able to change the rate at which applications are executed, and it cannot handle different priorities nor deadlines. For this reason, we have implemented InK in our event-based simulator and we have evaluated the behaviour of InK and our energy-aware scheduler. 
It is important to note that most of the parameters are chosen by the application developer or are device-specific, thus they are an input for our problem (i.e., they are not configurable by the algorithm). These parameters are the capacitor size ($C$), harvesting rate ($PH$), the voltage parameters ($V_{min}$, $V_{max}$, $V_{-1}$), tasks parameters (current consumption values, priority, execution time, deadline, order or arrival time). The only two parameters that can be configured by the algorithm or the event-based simulator are $\bigtriangleup t$ and the voltage turn-on threshold, $V_{th}$, which is a configurable parameter that corresponds to the voltage the capacitor needs to reach after a power failure in order to turn on again. While $\bigtriangleup t$ is only used by the optimization algorithm, $V_{th}$ is only used by Ink, as our energy-aware algorithm avoids power failures.

\begin{figure*}[t]
\centering
\begin{subfigure}{.85\textwidth}
\includegraphics[width=\linewidth]{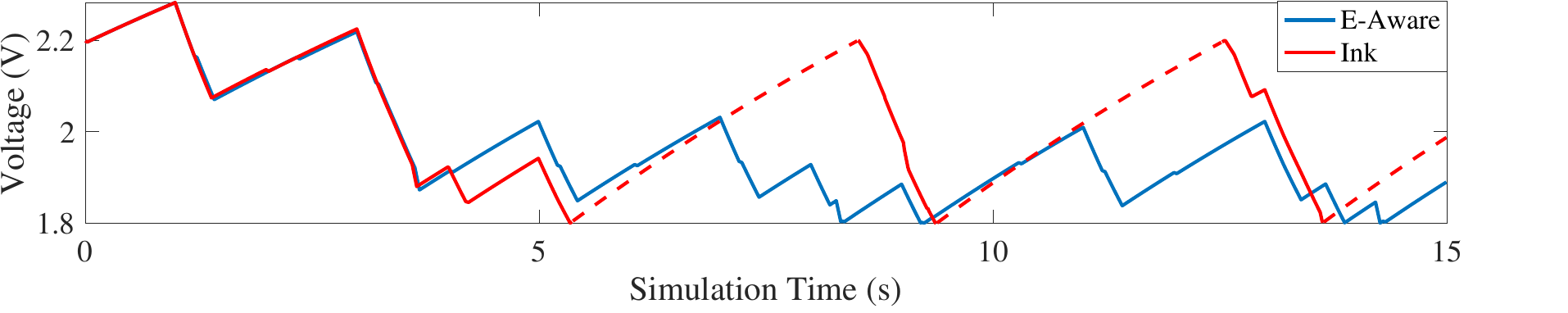}
\caption{Voltage variations when executing different tasks when $PH=5mW$}
\label{fig:V_32_V}
\end{subfigure}
\hfill
\begin{subfigure}{.85\textwidth}
\centering
\includegraphics[width=\linewidth]{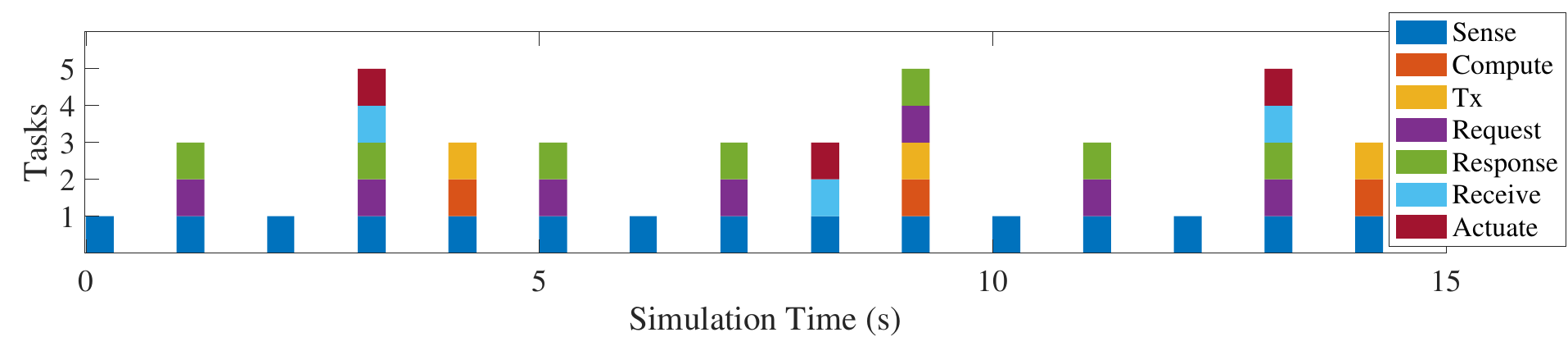}
\caption{Task arrival time }
\label{fig:V_32_Task}
\end{subfigure}
\hfill
\caption{Voltage variations for InK and our approach and tasks to be executed}
\label{fig:V_32}
\end{figure*}

Figure \ref{fig:V_32_V} shows the voltage across the capacitor for the two approaches (E-Aware and InK) when performing a sequence of tasks that need to be executed during the 15 seconds of the simulation and with a harvesting power of 5~mW. As can be seen in the Figure \ref{fig:V_32_Task} (where we show the arrival time of the tasks), Sensing happens every 1 second, while Request and Receive take place every 2 and 5 seconds, respectively. On the other side, Compute takes place after 5 Sense tasks. In this case, from the 41 tasks that need to be executed, InK is able to successfully schedule 21 tasks, while our energy-aware scheduler is able to schedule 36. While our solution is able to avoid power failures and never goes below 1.8V, InK has 3 power failures, at 5.37, 9.43 and 13.64 seconds. At these points in time, the device turns off (note the dashed line in Figure~\ref{fig:V_32}), consuming much less energy (we have considered it negligible) than in sleep mode (0.1 mA). However, when the voltage turn-on threshold is reached ($V_{th}=2.2V$), the device wakes up, which takes 0.1 seconds and consumes 3 mA. On the contrary, avoiding these power failures is beneficial to not waste energy in turning on and to not miss deadlines while in off.

\begin{figure}[t]
\begin{subfigure}{1\columnwidth}
\centering
\includegraphics[width=.88\columnwidth]{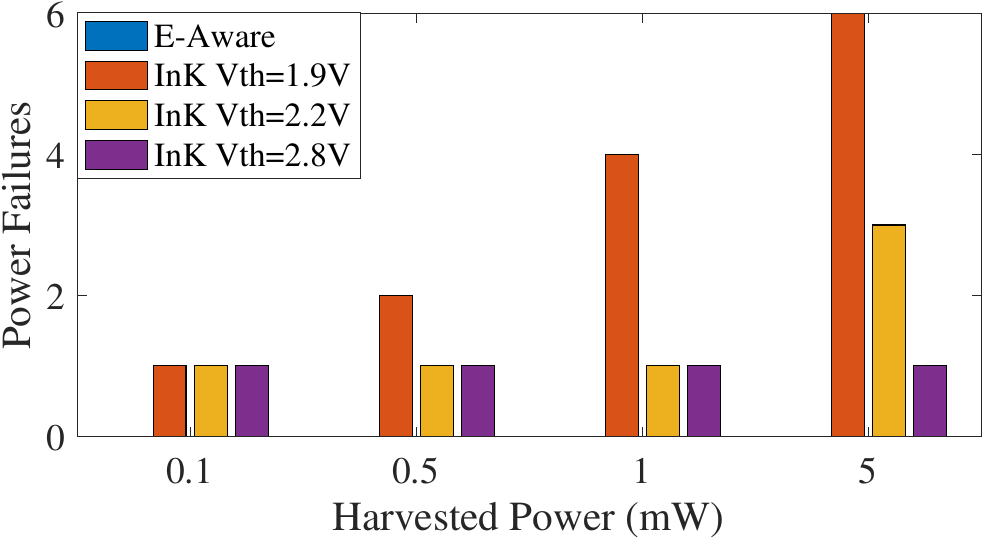}
\caption{Number of power failures}
\label{fig:PowerFailure}
\end{subfigure}
\hfill
\begin{subfigure}{1\columnwidth}
\centering
\includegraphics[width=.88\columnwidth]{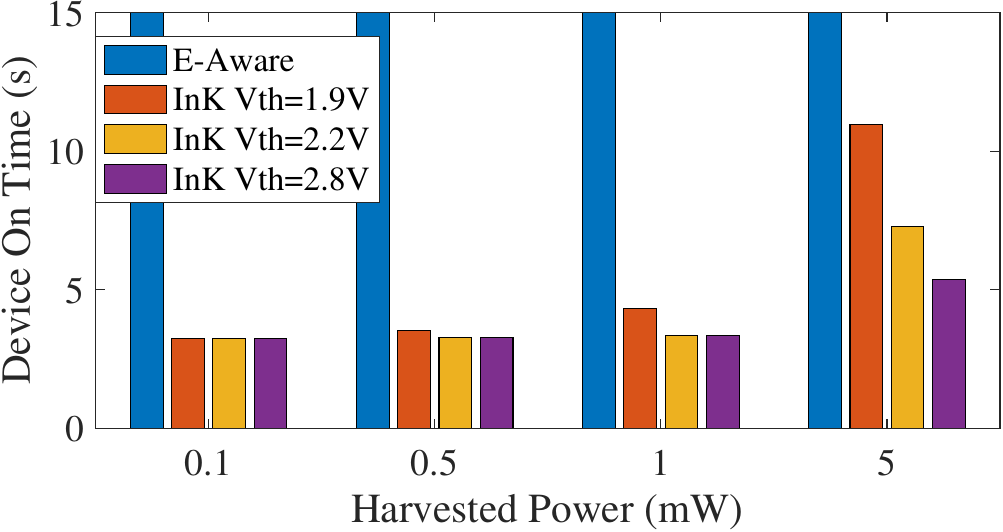}
\caption{Total time the device is On}
\label{fig:OnTime}
\end{subfigure}
\hfill
\caption{Comparison between the energy-aware scheduler and InK for different values of $PH$ when varying the turn-on threshold of InK}
\label{fig:pfailuresandTimeOn}
\end{figure}

\begin{figure}[t]
\begin{subfigure}{1\columnwidth}
\centering
\includegraphics[width=.88\columnwidth]{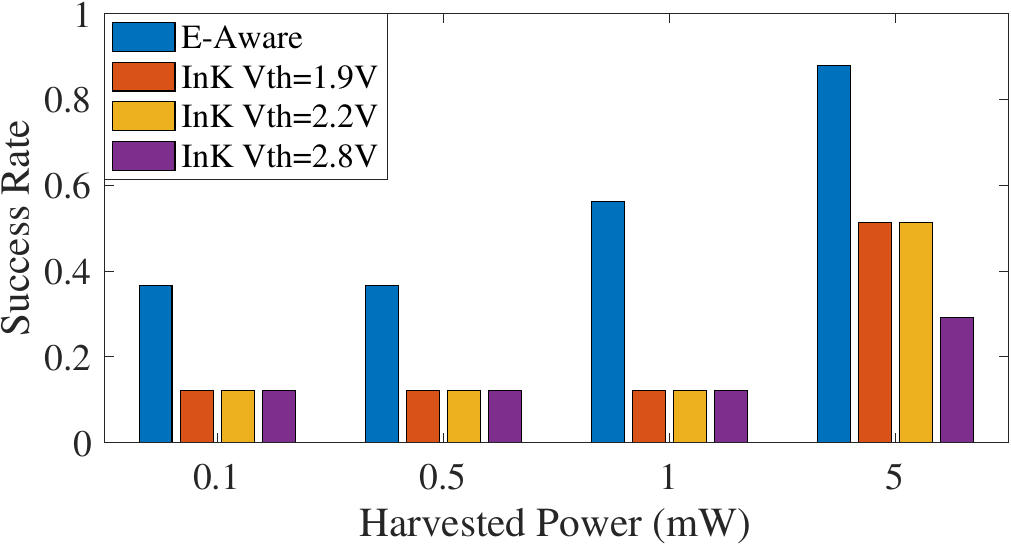}
\caption{Task Success Rate}
\label{fig:successRate}
\end{subfigure}
\hfill
\begin{subfigure}{1\columnwidth}
\centering
\includegraphics[width=.88\columnwidth]{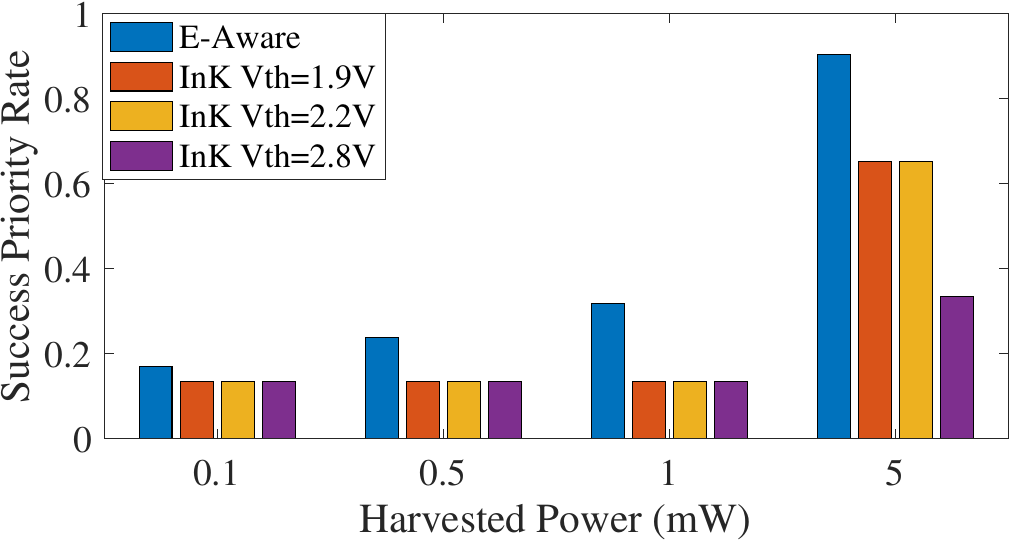}
\caption{Priority Success Rate}
\label{fig:successRatePrio}
\end{subfigure}
\hfill
\caption{Performance comparison between the energy-aware scheduler and InK for different values of $PH$ and a 4.7mF Capacitance when varying the turn-on threshold of InK}
\label{fig:SuccessRate}
\end{figure}

To give a better overview of the improvement of our energy-aware solution, in Figures \ref{fig:pfailuresandTimeOn} and \ref{fig:SuccessRate} we show the comparison between our energy-aware task scheduler and InK. We have considered the experiment setup explained in Section \ref{sec:methodology} with $PH$ of 0.1mW, 0.5mW, 1mW and 5mW, which are in line with the energy that can be obtained from indoor light \cite{Shirvanimoghaddam2018}. 
We show different values of $V_{th}$ for Ink but we do not need to define any turn on threshold for our energy-aware formulation as we ensure that the device will not turn off.

First, in Figure \ref{fig:PowerFailure} we see how many power failures occur in both approaches. And as expected, our energy-aware scheduler avoids power failures in all the cases, while InK is not able to manage them as it is not aware of the energy. 
This is possible thanks to the fact that our approach is  able to completely avoid power failures only if we assume perfect knowledge on the energy harvested (as it is the case in this specific experiment). However, if this knowledge is not perfect, some failures would occur. In order to provide a fair comparison, we have considered the effect of different values of the turn-on threshold voltage ($V_{th}$) for InK, and as we have seen that higher values than 2.8V do not provide better results, we only consider 1.9V (close to $V_{min}$), 2.2V and 2.8V.

For Ink we see that increasing the harvested power (from 0.1mW up to 5mW), for the lowest turn-on voltage threshold ($V_{th}=1.9V$) the number of power failures increases, while for the case of 2.8V of threshold, there is only 1 power failure. However, and as can be seen in Figure \ref{fig:OnTime}, where we show the total time the device is On, we can see that the device is only awake for 3.5 seconds till $1mW$ (in which InK is able to only schedule 5 tasks) and that in the case of 5mW it stays on 5.34 seconds. However, this difference is due to the fact that the device starts at 2.2V and the $PH$ is enough to be awake for more than 5 seconds, but after that, it is not possible to reach the 2.8V to turn on again during the 15 seconds of the simulation. This is mainly due to the fact that the harvested energy is too low to charge the capacitor up to the voltage turn-on threshold. On the other side, and as mentioned, the low energy harvesting rate is not an issue for our energy-aware solution and the device remains On all the time.

In Figure \ref{fig:successRate} we show the task success rate, and the priority success rate is shown in Figure \ref{fig:successRatePrio}. While the task success rate only represents the rate of the tasks that can get scheduled from the total number of tasks that need to be scheduled, success priority rate is the success rate multiplied by all the task priorities.
The simulation time for our experiments has been set to 15 seconds, where we assume that we have total knowledge on all the tasks that will be scheduled and the power that will be harvested during the entire simulation period. 

For all the values of harvested power considered, our E-aware approach provides better results, no matter the turn-on threshold voltage value chosen by InK. However, the lower the harvested power is, the less improvement we are able to see, as many of the tasks require more energy than what is available in the capacitor. Since the $2.8V$ turn-on threshold provides the worst success (priority) rate, and as the lower one ($1.9V$) does not provide better results for InK and can lead to more power failures, in the rest of the paper we have considered $V_{th}=2.2V$.

\begin{figure}[t]
\centerline{\includegraphics[width=0.85\columnwidth]{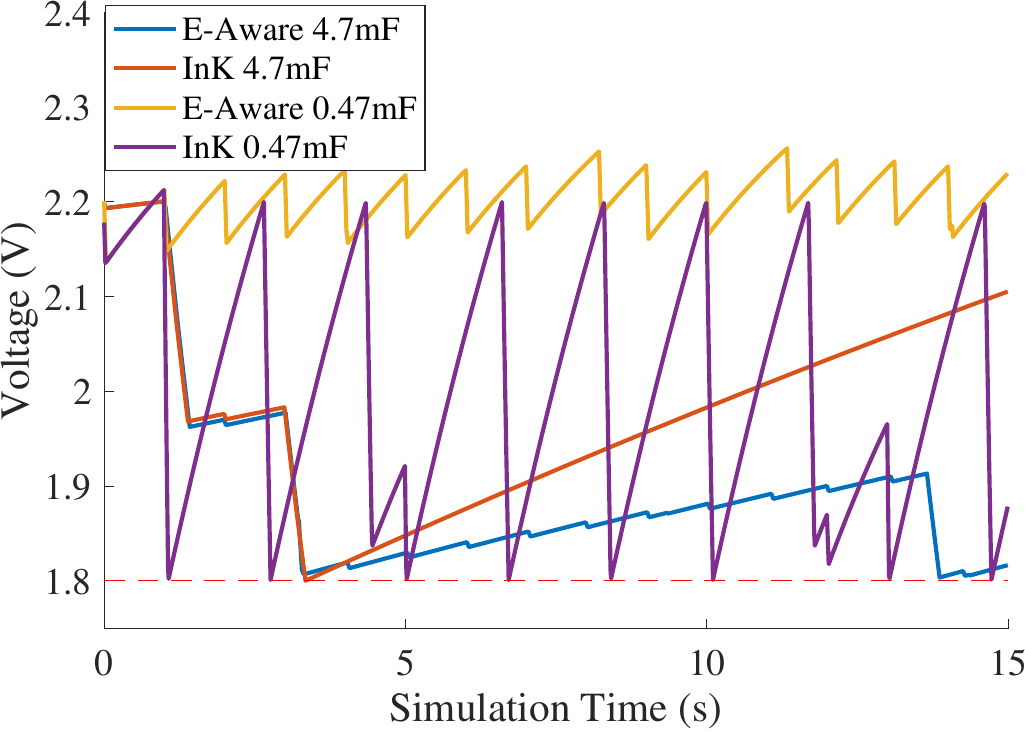}}
\caption{Voltage variations when executing different tasks for 1mW of $PH$}
\label{fig:V_3758}
\end{figure}

In order to reduce the time the capacitor needs to reach the voltage turn-on threshold, we have also considered a smaller capacitor. In Figure \ref{fig:V_3758} we show the voltage variations over time when executing some tasks for a capacitor of 4.7 mF and 0.47 mF when harvesting 1mW of power. For the capacitor of 4.7 mF it takes longer to charge, and as can be seen, the energy-aware solution keeps increasing its voltage while executing non power hungry applications until time instant 13.66 seconds, where it is able to execute a more powerful task. However, InK turns off at time instant 3.35 seconds and it is not able to turn on again. When using a smaller capacitor of 0.47 mF a complete different behaviour is seen. Our energy-aware solution is able to schedule 18 applications, although all of them have low priority, resulting in a total profit of 24. In contrast, InK keeps turning off and on. But the problem is that with this low energy harvesting power and small capacitor, the energy needed to turn on and just execute an immediate and powerful task is not enough. Since InK is not aware of the energy, it chooses the task with the highest priority which fulfills the deadline constraints, without worrying about its energy consumption. The problem is that while trying to execute the task, its voltage drops to $V_{min}$ and the device turns off. After reaching the turn-on threshold it turns on again and tries to re-execute the same application if the deadline is not reached yet. In this way, InK is only able to successfully execute 2 tasks.  

\begin{figure}[t]
\centerline{\includegraphics[width=0.85\columnwidth]{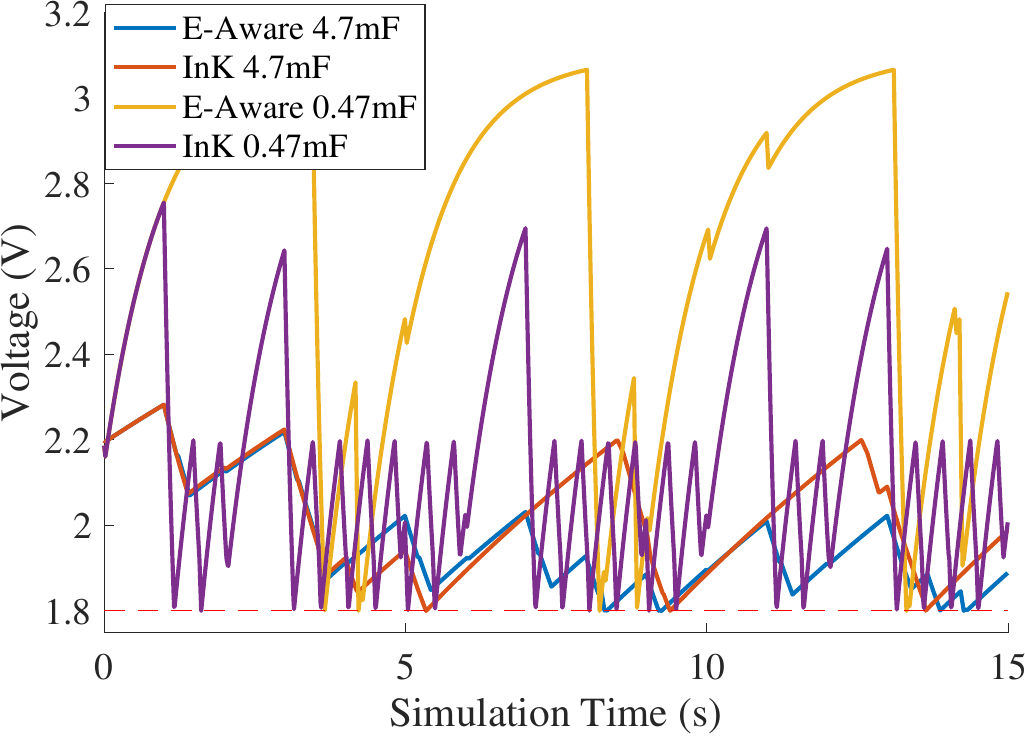}}
\caption{Voltage variations when executing different tasks for 5mW of $PH$}
\label{fig:V_3253}
\end{figure}

In Figure \ref{fig:V_3253} we increase the harvested power to 5mW (i.e., $PH=5mW$), so more tasks are scheduled. For the smaller capacitor ($C=0.47mF$) we see how it gets charged faster, and in the energy-aware scheduler, we see how it intelligently decides to get almost fully charged to have enough energy to be able to execute a more energy-hungry task in the time instants 3.4, 7.96 and 13.1 seconds. In the meantime, if it can achieve a higher reward by executing a less-power hungry task during the charging process, it will do it.
While InK is only able to schedule 7 tasks, our energy-aware solution schedules 18 tasks. For the 4.7 mF capacitor, we can observe that it takes longer to charge the capacitor, but also the voltage drop is less abrupt. In this case, while our energy-aware solution it is able to execute 36 tasks, InK is only able to execute 21. 

\subsection{Influence of look-ahead window on performance}

In the previous section, we showed the potential of energy-aware scheduling, by evaluating the maximum gain in performance when assuming perfect prediction of all future tasks and energy harvesting power. In this section, we evaluate how the size of the look-ahead window influences the effective performance gain in a real system (as perfect prediction over an infinite window is not achievable in practice). In batteryless devices, energy consumption is the main parameter to be aware off, and being able to predict both the available energy and the energy to be consumed will allow to better schedule the tasks.
However, knowing how much energy is expected in the future gets harder the further in time it is predicted. And also, more memory is needed to perform those calculations. For this reason, it is important to know how far in the future the scheduler needs to look to achieve the best possible improvement.

\begin{figure}[t]
\begin{subfigure}{1\columnwidth}
\centering
\includegraphics[width=0.88\columnwidth]{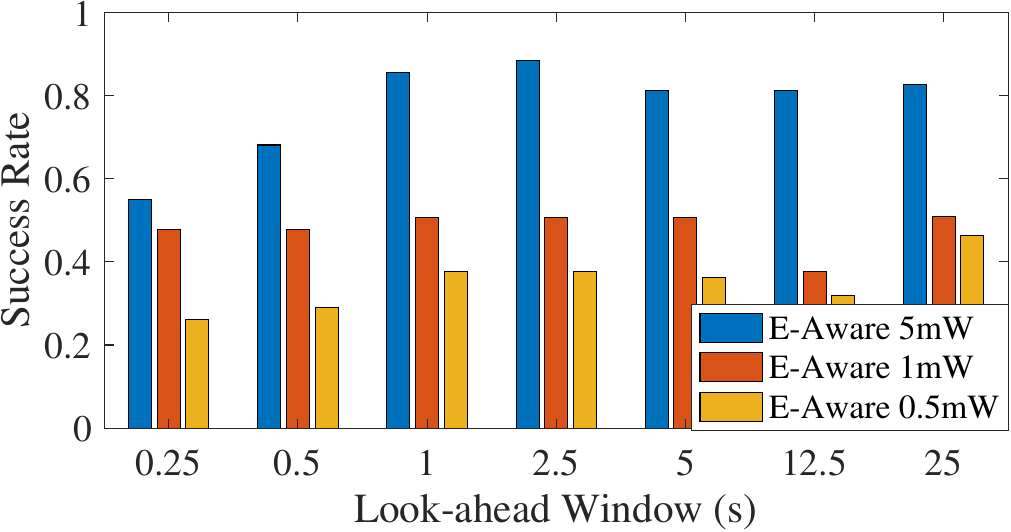}
\caption{Task Success Rate}
\label{fig:SuccessWindow_2.2}
\end{subfigure}
\hfill
\begin{subfigure}{1\columnwidth}
\centering
\includegraphics[width=0.88\columnwidth]{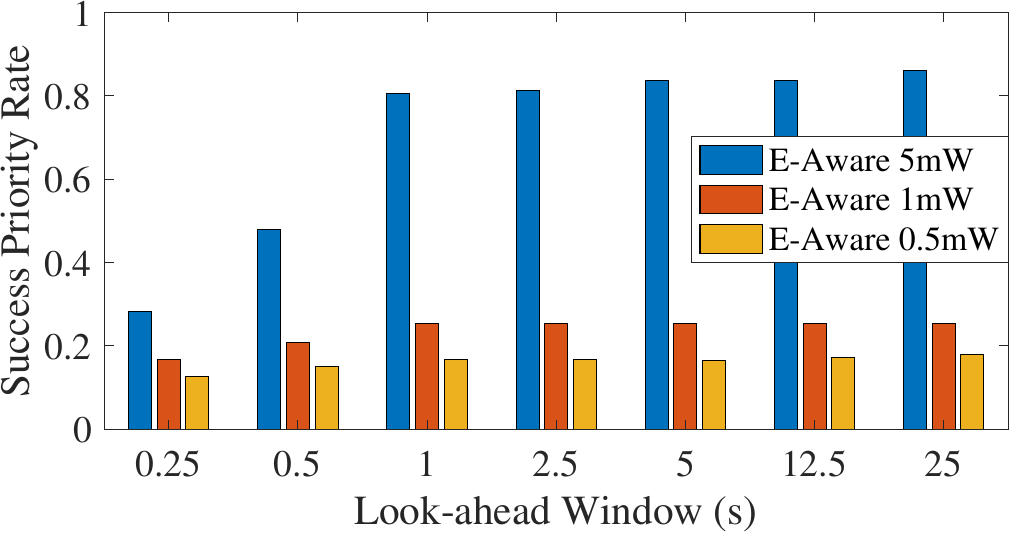}
\caption{Priority Success Rate}
\label{fig:SuccessPrioWindow_2.2}
\end{subfigure}
\hfill
\caption{Performance of our energy-aware scheduler when varying the look-ahead window time for $V_{-1}=2.2V$}
\label{fig:successTime2.2}
\end{figure}

For this reason, in Figure \ref{fig:successTime2.2} we show the task success rate and the priority success rate of a simulation of 25 seconds when we let the scheduler look ahead only over a limited look-ahead optimization window and re-execute it at the start of each window. 
We define small look-ahead optimization windows with different sizes starting from 0.25 seconds (meaning 100 look-ahead optimization windows) up to 25 seconds, which means one look-ahead optimization window. 
As expected, when increasing the look-ahead window time a better performance in terms of success priority rate is achieved. 
However, in terms of tasks, it seems that more tasks are executed if we optimize every 2.5 seconds (c.f., Figure \ref{fig:SuccessWindow_2.2}), but since we are optimizing the priority of these scheduled tasks, in Figure \ref{fig:SuccessPrioWindow_2.2} we see that the profit is not significantly improved after 1 second of look-ahead window time. 
This means that when increasing the look-ahead window, less tasks are executed, but that these tasks have a higher priority. 
For this reason, if the aim is to only maximize the number of tasks, a look-ahead window size of 1 second will be enough. 
However, only looking at a very short look-ahead window, tasks that consume a lot of power will never be executed. In our experiments, these tasks are also the tasks that have a high priority, and for this reason if more priority tasks need to be executed, a 5 seconds look-ahead window size will be needed.

\begin{figure}[t]
\begin{subfigure}{1\columnwidth}
\centering
\includegraphics[width=0.88\columnwidth]{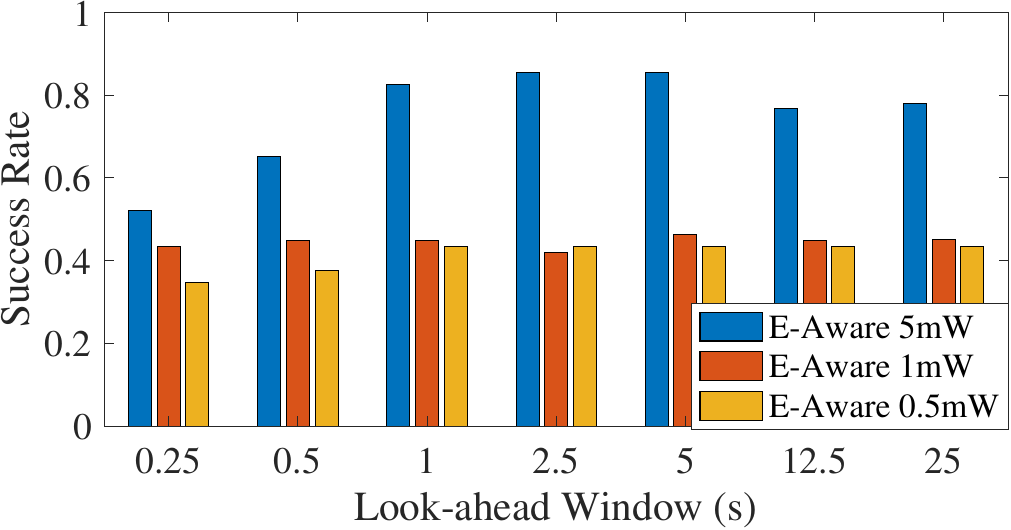}
\caption{Task Success Rate}
\label{fig:SuccessWindow_1.9}
\end{subfigure}
\hfill
\begin{subfigure}{1\columnwidth}
\centering
\includegraphics[width=0.88\columnwidth]{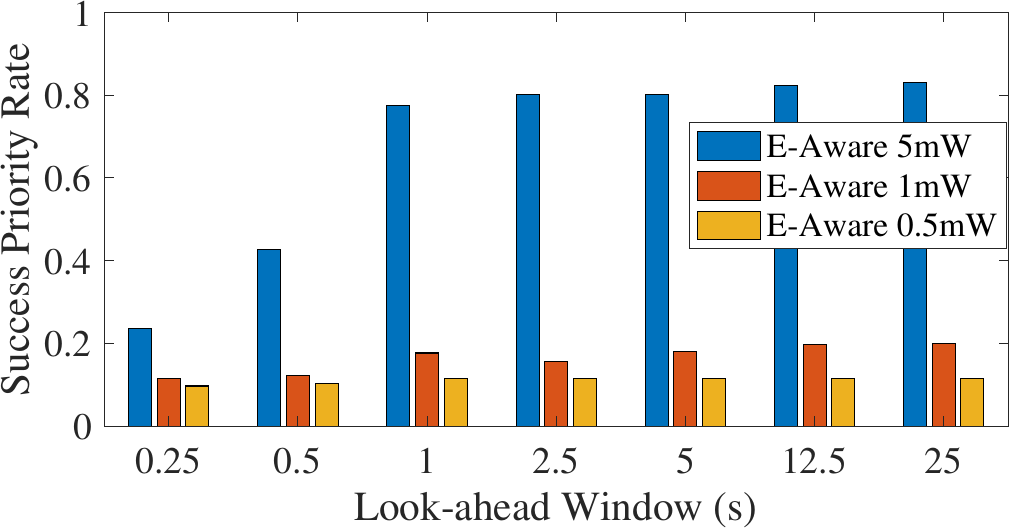}
\caption{Priority Success Rate}
\label{fig:SuccessPrioWindow_1.9}
\end{subfigure}
\hfill
\caption{Performance of our energy-aware scheduler when varying the look-ahead window time for $V_{-1}=1.9V$}
\label{fig:successTime1.9}
\end{figure}

In general terms, and for the considered task set, if we look at the look-ahead time window and the number of tasks executed per second, we can say that a look-ahead window that can look ahead up to 4 tasks will lead to a good performance in terms of successful task executions, but if more power hungry tasks need to be deployed, looking ahead 8 tasks in the future is needed. To make sure that our conclusions can be generalized to other initial capacitor voltages $V_{-1}$, we also show the results when the initial voltage of the capacitor is $V_{-1}=1.9V$ in Figure \ref{fig:successTime1.9}. The same conclusions are still valid, although a performance reduction of $16\%$ occurs for short look-ahead windows (0.25 seconds). However, when increasing the look-ahead window size to 12.5 seconds, the difference is only $0.016\%$. This means that the bigger the look-ahead window time is, the lower impact the initial voltage has, but also longer experiment duration will diminish this effect.

\section{Conclusions and Future Work}\label{sec:conclusions}

In this paper we have shown that energy-aware scheduling mechanisms are needed to improve the performance of successful application execution on batteryless devices. These tiny devices frequently turn on and off, and being aware of the expected energy consumed and the energy that can be harvested is crucial. 
For this reason, in this paper, we provide theoretical insights into the achievable performance gain of energy-aware task scheduling, compared to state of the art non-aware batteryless application task schedulers. Moreover, we study the influence of the size of the look-ahead energy prediction window, as a first step towards developing a practical scheduling heuristic that can run on batteryless devices.
To do so, we have proposed a new optimal energy-aware scheduling algorithm that takes into account the energy available in the capacitor and the expected energy to be harvested to optimally schedule the tasks, which are defined by their priority, arrival time, execution time, energy consumption, and set of task parents that need to be executed beforehand. We have compared our energy-aware solution against InK, an energy-unaware dynamic scheduler based on priorities and deadlines. Our results show that making the task scheduler energy aware avoids power failures, which allows more tasks to make their deadlines. 
And finally, we have evaluated how much look-ahead window time in the future is needed to achieve optimal performance, and we can conclude optimizing every 4 tasks will optimize the task scheduled rate, but power-hungry tasks will suffer from it. In fact, increasing the number of optimized tasks up to 8 tasks will help to obtain a better successful rate where all kind of tasks can be scheduled.

There are several future research directions.
The results presented in this work can now be used as a basis for heuristic schedulers that can be executed in real-time on batteryless devices, and can also be used to define the requirements for energy harvesting and consumption prediction techniques for such schedulers.
However, there are also several challenges and difficulties that should be taken into account when applying the insights of this work in real life. Firstly, distilling from the decisions made by the MILP, more straight forward rules that are able to select a task with very limited look-ahead calculations should be considered. Secondly, and in order to make decisions, the algorithm needs the inputs described in this paper. However, obtaining all this necessary information as inputs could require some more advanced circuitry (i.e., obtaining the capacitor voltage or the harvesting power), whose energy consumption should also be taken into account. And thirdly, a real implementation needs to consider (and minimize) the energy consumed by the scheduler itself as well, but also other factors such as the operating systems underlying effects. 

In order to tackle these challenges, first, we should design a more light-computing suboptimal solution (i.e., heuristic approach), to be solved in these energy constrained devices. There are different and straight forward solutions (e.g., greedy approaches, genetic algorithms, simulated annealing) that need to be investigated to determine which one offers better performance. Second, a dedicated circuitry module should be added to the design. This additional module should efficiently read the voltage of the capacitor with a resistor divider, but also should be able to obtain the harvested power. This would depend on the type of harvester to be used. For example, if a solar panel is used, the circuitry should also be able to read the open circuit voltage and the short circuit current of the photovoltaic cell, which should be done periodically. To read these values, we need to disconnect the photovoltaic cell from the capacitor, incurring in some waste of energy that should also be considered when addressing the harvested power. However, as an alternative to doing measurements, the use of prediction methods could be used. These techniques normally depend upon statistical and stochastic models of harvested energy using linear regression, Exponential Weighted Moving Average, Markov chains or machine learning. Depending on the look-ahead window, short-time prediction (but also low complexity) techniques should be further investigated. 

Finally, aiming at considering the energy consumption of this circuitry, the scheduler and other factors such as the effects of the underlying operating system, new energy measurements would need to be taken. These measurements should be performed in a huge variety of conditions and environments to have a broader view of the expected energy consumption.


%



\ifCLASSOPTIONcompsoc
  \section*{Acknowledgments}
\else
  \section*{Acknowledgment}
\fi

Part of this research was funded by the Flemish FWO SBO S004017N IDEAL-IoT (Intelligent DEnse and Long range IoT networks) project, the University of Antwerp IOF funded project COMBAT (Time-Sensitive Computing on batteryless IoT Devices), the Flemish FWO SBO S001521N IoBaLeT (Sustainable Internet of batteryless Things) project and the CERCA Programme, by the Generalitat de Catalunya. The computational resources and services used in this work were provided by the VSC (Flemish Supercomputer Center), funded by FWO and the Flemish Government - department EWI.

\ifCLASSOPTIONcaptionsoff
  \newpage
\fi



%


\bibliographystyle{IEEEtran}
\bibliography{./biblio}

%


\begin{IEEEbiography}[{\includegraphics[width=1in]{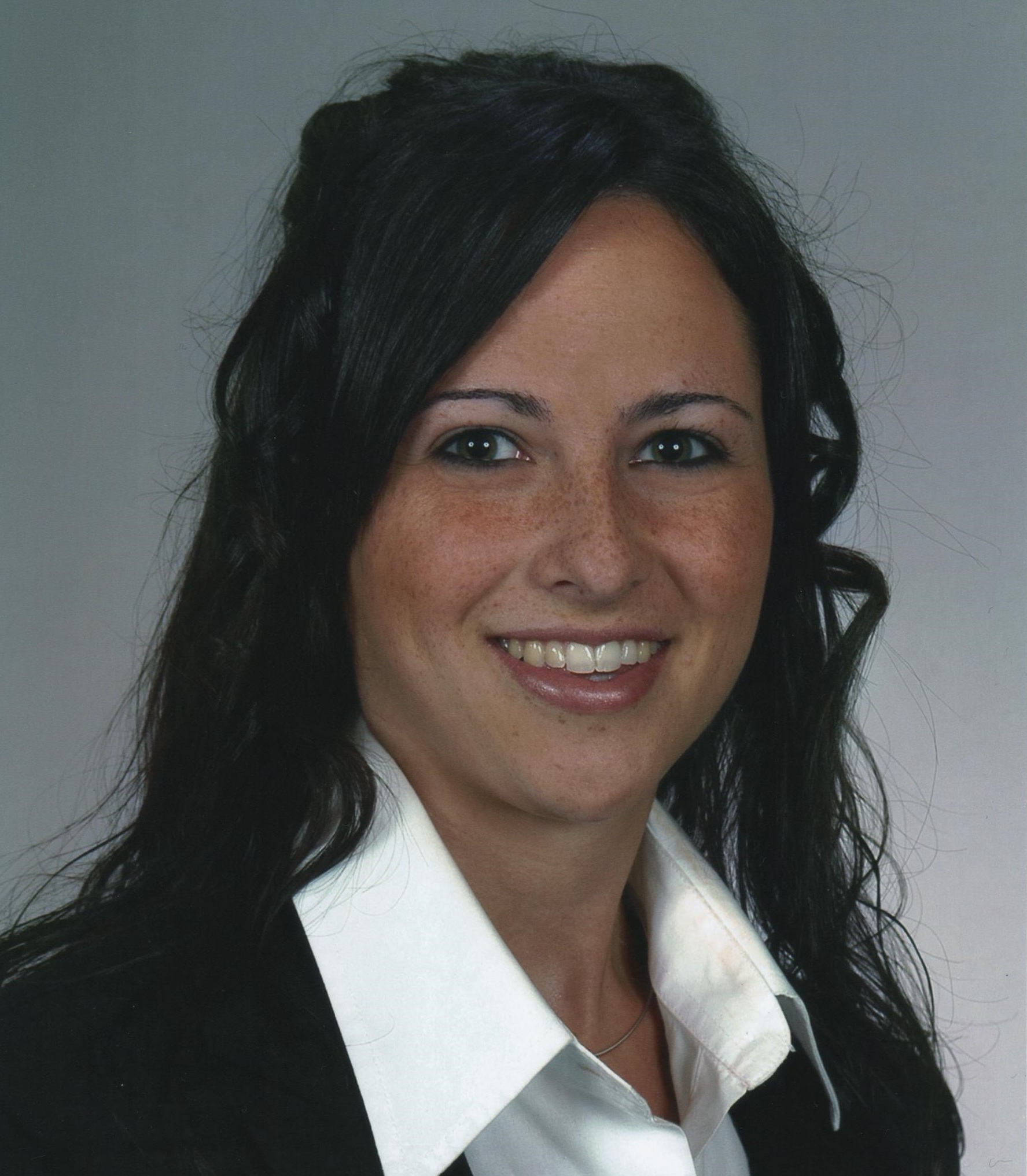}}]{Carmen Delgado}
She received the M.Sc. in telecommunications engineering, the M.Sc. degree in biomedical engineering and a Ph.D. (cum laude) in Mobile Network Information and Communication Technologies from the University of Zaragoza, Spain, in 2013, 2014, and 2018 respectively. 
She joined the Internet Technology and Data Science Lab (IDLab) of the University of Antwerp, associated with imec, Belgium as a post-doctoral researcher in 2018.
She is currently working in the i2CAT Foundation as senior researcher. Her research interests lie in the field of Internet of Things, resource allocation, energy harvesting, low power communications, energy modeling and performance evaluation of wireless sensor networks.
\end{IEEEbiography}

\begin{IEEEbiography}[{\includegraphics[width=1in]{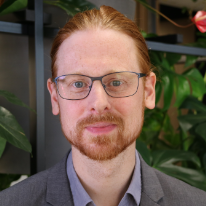}}]{Jeroen Famaey}
He is an assistant professor associated with imec and the University of Antwerp, Belgium. He received his M.Sc. degree in Computer Science from Ghent University, Belgium in 2007 and a Ph.D. in Computer Science Engineering from the same university in 2012. He is co-author of over 120 articles published in international peer-reviewed journals and conference proceedings, and 10 submitted patent applications. His research focuses on performance modeling and optimization of wireless networks, with a specific interest in low-power, dense and heterogeneous networks.
\end{IEEEbiography}









\end{document}